\documentclass[%
 reprint,
nofootinbib,
nobibnotes,
bibnotes,graphicx,graphics,
 amsmath,amssymb,
 aps,stmaryrd,
]{revtex4-1}
\usepackage{natbib} 
\usepackage{soul}
\newcommand{\comment}[1]{}
\usepackage{tikz}
  \newlength\squareheight
\usepackage{wasysym} 
\usepackage{graphicx}
\usepackage{dcolumn}
\usepackage{bm}

\begin{document}

\preprint{Draft}

\title{Symbiotic and antagonistic disease dynamics on networks using bond percolation}

\author{Peter Mann}
\email{pm78@st-andrews.ac.uk}
\author{V. Anne Smith}%
\author{John B.O. Mitchell}
\author{Simon Dobson}
\affiliation{School of Computer Science, University of St Andrews, St Andrews, Fife KY16 9SX, United Kingdom }
\affiliation{School of Chemistry, University of St Andrews, St Andrews, Fife KY16 9ST, United Kingdom }
\affiliation{School of Biology, University of St Andrews, St Andrews, Fife KY16 9TH, United Kingdom }

\date{\today}

\begin{abstract}
In this paper we introduce a novel description of the equilibrium state of a bond percolation process on random graphs using the exact method of generating functions. This allows us to find the expected size of the giant connected component (GCC) of two sequential bond percolation processes in which the bond occupancy probability of the second process is modulated (increased or decreased) by a node being inside or outside of the GCC created by the first process. In the context of epidemic spreading this amounts to both a antagonistic partial immunity or a synergistic partial coinfection interaction between the two sequential diseases. We examine configuration model networks with tunable clustering. We find that the emergent evolutionary behaviour of the second strain is highly dependent on the details of the coupling between the strains. Contact clustering generally reduces the outbreak size of the second strain relative to unclustered topologies; however, positive assortativity induced by clustered contacts inverts this conclusion for highly transmissible disease dynamics.  
\end{abstract}

\pacs{Valid PACS appear here}
\maketitle



A network is a collection of nodes that are connected by edges. Bond percolation on complex networks is a widely studied binary-state stochastic process where the edges of a network are said to be \textit{occupied} with probability $T_1$ or \textit{unoccupied} with probability $1-T_1$. When $T_1$ is small, many edges are unoccupied and the network is fractured into small components. At some critical value, $T_{1,c}$, the small components connect together to form a macroscopic giant connected component (GCC); the expected size of the GCC exhibits a second-order phase transition. As $T_1\rightarrow 1$ the GCC occupies an increasing fraction of the network. Nodes not contained within the GCC are said to be in the residual graph (RG) of the percolation process. 

The size of the GCC following bond percolation is equivalent to the fraction of nodes that are in the removed state of the SIR process. In the SIR model, nodes are either susceptible, infected or removed. Infection occurs along edges that connect infected nodes to their susceptible neighbours. Once infected, a node remains infected for a period, $\tau$, before recovering to the $R$ state. The equivalence between SIR dynamics and bond percolation occurs when the infection period is a single-valued distribution \cite{PhysRevE.76.036113}. Hence, the absorbing equilibrium of the model is binary-state and is composed of only susceptible and recovered nodes.

\begin{figure}[ht!]
\begin{center}
\includegraphics[width=0.48\textwidth]{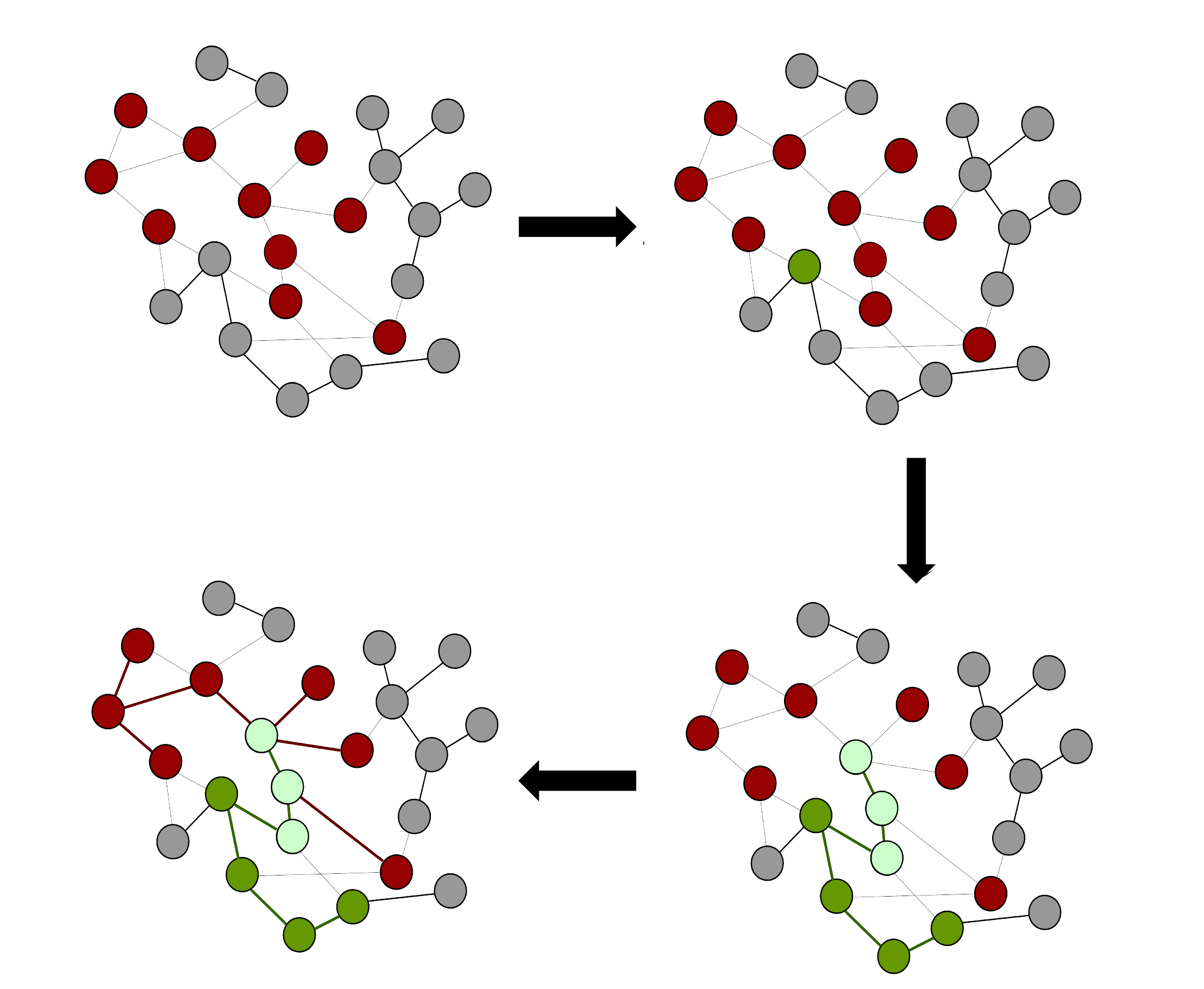}
\caption[networkfollowing1]{(From top left) A primary disease has spread over a network of susceptible nodes (grey) to create a giant component of infected hosts (red) at its equilibrium. A seed node (dark green) of degree $k=4$ has degree 2 in the RG (uninfected neighbours) and $k-l=2$ strain 1 infected neighbours. In the general case, the second strain then spreads on \textit{both} the RG with transmissibility $T_2$ \textit{and} over the GCC of strain 1 (light green) with modulated transmissibility $T_2'$. The light green nodes are \textit{coinfected} with both strain 1 and strain 2, while the red (dark green) nodes only have strain 1 (2).
} \label{fig.network1}
\end{center}
\end{figure}

Once a network has been percolated, it can be percolated a second time. Under the disease mapping, this corresponds to a second disease, temporally separated from the first, spreading over the network. This could be a mutant strain of strain 1 invading a population, or a different pathogen altogether, we will simply refer to `strain 2'. If strain 2 does not interact with strain 1 in any way, then the expected dynamics will follow the traditional SIR model and ordinary bond percolation is sufficient to describe the outbreak fraction. However, strain 1 could grant perfect cross-immunity to recovered nodes such that they do not become infected further \cite{newman_2005,PhysRevE.84.036106,PhysRevE.81.036118,PhysRevE.84.026105,PhysRevE.87.060801,PhysRevLett.110.108103,PhysRevE.103.062308}. Alternatively, infection by strain 1 could be a prerequisite to contracting strain 2 \cite{10.1371/journal.pone.0071321,PhysRevE.96.022301,2020arXiv201209457M}. In both cases, the \textit{infection history} of a node in the equilibrium of strain 1 is important in whether or not it contracts the second strain. The more likely scenario, and the most general disease interaction model, is that the spreading of strain 2 is only \textit{modulated} by the presence of strain 1. The modulation could facilitate strain 2, a \textit{partial coinfection} model, or it could hinder its spreading, a \textit{partial immunity model}. With these definitions in place, the above two scenarios of perfect cross-immunity and perfect coinfection are the limiting logic of a smooth spectrum of interactions in which only partial interaction is observed and it is this case that we model in this paper. We illustrate the concepts of partial coinfection
and partial immunity in Figure \ref{fig.network1}.

We choose to investigate our model for clustered networks; that is, networks containing a non-zero density of triangles. We do this because typically, human contact networks are highly clustered: any two contacts often have a third, mutual friend; and, clustering has important consequences for real epidemics \cite{watts_strogatz_1998,rothenberg_potterat_woodhouse_muth_darrow_klovdahl_1998,szendroi_csanyi_2004,read_keeling_2003}. Clustering has been well studied in the literature and is known to greatly influence the properties of the emergence and overall size of the GCC of strain 1. It is not only the number of triangles in a network that determines the percolation properties, but also the nature of how those triangles connect together and to other edge topologies; degree assortativity being a leading factor. Therefore, different clustered degree distributions often lead to dichotomous conclusions regarding the effect of clustering. It was shown by Newman \cite{PhysRevLett.103.058701} that Poisson networks with clustering reduce the epidemic threshold of the first strain, whilst Miller \cite{miller_2009,citekey2} and Gleeson \cite{PhysRevE.81.066114,PhysRevE.83.056107,PhysRevE.91.052807} showed that clustering can increase the epidemic threshold when assortativity is controlled, which was also reported by Hasegawa \textit{et al} \cite{PhysRevE.101.062310,HASEGAWA2021125970}. 

Recent results \cite{PhysRevE.103.062308,2020arXiv201209457M} have investigated the role of clustering for a second strain in the limiting cases of perfect cross-immunity (spreading solely on the RG) and perfect coinfection (spreading solely on the GCC) using bond percolation. It was found that clustering can again exhibit polarised results depending on the details of the contact network. In this paper, we introduce a model of partial immunity using the generating function formulation and find excellent agreement with Monte Carlo simulations of bond percolation. This model generalises the results of \cite{PhysRevE.103.062308,2020arXiv201209457M} by relaxing the strict conditions imposed on how the second strain interacts with the first one (spreading on both the RG and the GCC). We then study the role of clustering in shaping the evolutionary behaviour of strain 2 for power law networks with tunable clustering that exhibit exponential degree cut-offs.

\section{Configuration model and generating functions}

The configuration model is a prescription to create random graphs whose degrees are distributed according to some predefined distribution \cite{PhysRevLett.85.5468, PhysRevE.64.026118,PhysRevE.66.016128,fosdick_larremore_nishimura_ugander_2018,newman_2019}. The central object of the model is the degree distribution, $p(k)$, which is the probability that a node chosen at random from the network has degree $k$. By extending $p(k)$ to a joint distribution, $p(s,t)$, where $s$ is the number of ordinary tree-like edges and $t$ is the number of triangles that the node is involved in (such that $k=s+2t$), the configuration model can be used to create random graphs with clustering \cite{PhysRevLett.103.058701, miller_2009}. Similarly, by extending the joint distribution to include other cycles, such as squares, 4-cliques and so on, graphs with complicated clustering can be created.

To create a realisation of a clustered graph of size $N$ according to the configuration model, each node is assigned half-degrees or \textit{stubs}, in other words, an $(s_i,t_i)$ pair drawn from the joint distribution $p(s,t)$ where the following conditions hold
\begin{equation}
    \sum_{i=1}^N s_i = 2\mathcal S
\end{equation}
and 
\begin{equation}
    2\sum_{i=1}^N t_i = 3\mathcal T
\end{equation}
where $\mathcal S$ is the number of tree-like edges in the network and $\mathcal T$ is the number of triangles, both of which are integers, to ensure that the drawn sequence is graphic. 
At each step of the construction, pairs of tree-like stubs from two different randomly selected nodes $i$ and $j$ are connected together to create a full edge; similarly, three triangle stubs are selected at random from three different nodes $i,j,k$ to create a triangle among them. This process is repeated until all stubs have been matched among the nodes. Each constructed graph is a member of an ensemble of graphs with equivalent degree distributions and is absent of degree correlations \cite{newman_2019}. 

Next we turn our attention to the generating function formulation \cite{PhysRevLett.85.5468,PhysRevE.64.026118,PhysRevLett.103.058701,miller_2009}. Generating functions are infinite series that can be used to encapsulate probabilities associated with certain network properties. For instance, the joint probability distribution is generated by a bivariate generating function as
\begin{equation}
    G_0(z_\bot,z_\Delta) = \sum^\infty_{s =0}\sum^\infty_{t=0} p(s ,t){z_\bot}^{s }{z_\Delta}^{t}\label{eq:G0_generic}
\end{equation}
This expression is understood as follows: we choose a node at random from the network and let $z_\bot$ and $z_\Delta$ be the probabilities that a single tree-like edge and a single triangle is in state $Z_\bot$ and $Z_\Delta$, respectively. Assuming that all tree-like edges and all triangles are independent of one another (such that $Z_\bot$ and $Z_\Delta$ are independent and identically distributed probabilities), then the probability that precisely $s$ edges and $t$ triangles surrounding a node are in this state is $z_\bot^s$ and $z_\Delta^t$. We then multiply this expression by the probability that the node we chose had joint degree $(s,t)$, which is simply $p(s,t)$ and finally, we sum over all combinations of $s$ and $t$ in the network. This yields the probability of choosing a node at random, or equivalently, the fraction of nodes in the network whose tree-like edges are in states $Z_\bot$ and whose triangles are in state $Z_\Delta$.

If $z_\bot=z_\Delta=1$, then these states occur with certainty, we have a normalising condition
\begin{equation}
    \sum^\infty_{s =0}\sum^\infty_{t=0} p(s ,t) = 1
\end{equation} 
The probability of choosing a node at random from the network and the probability of choosing an \textit{edge} at random and
then picking one of the adjacent nodes at random do not yield equivalent probabilities in general. This is because high degree nodes are more likely to be chosen and so, the properties of the nodes are different to one another. Further, the properties of the nodes reached by following a random tree-like edge and a random triangle are also non-equivalent in general. Thus, we must write an expression for the probability of reaching a node of joint degree $(s ,t)$ by traversing a random tree-like edge back to a node
\begin{equation}
    G_{1,\bot}(z_\bot,z_\Delta) = \frac{1}{\langle s  \rangle}\frac{\partial G_0}{\partial z_\bot},
\end{equation}
and by following a random triangle back to a node
\begin{equation}
    G_{1,\Delta}(z_\bot,z_\Delta) = \frac{1}{\langle t \rangle}\frac{\partial G_0}{\partial z_\Delta}
\end{equation}
In each case, $\langle s\rangle$ is the average number of tree-like edges a node is a member of and is given by $\partial_{z_\bot} G_0(1,1)$ with an analogous expression for $\langle t\rangle$. 

The bond percolation equivalence can now be understood in detail. If we set $z_\bot$ as the probability that a single tree-like edge does not connect the chosen node to the GCC, and $z_\Delta$ as the probability that the chosen node's involvement in a single triangle fails to connect it to the GCC, then we can generate the probability that a randomly chosen node fails to be part of the GCC. Thus, we can find emergent macroscopic properties of the entire network from the description of the local environment of a node, that we then average over all permissible node joint degrees.

\section{Partial interaction}

We begin this section with a review of traditional generating function theory for tree-like edges \cite{PhysRevE.66.016128}. To obtain the faction of the network that did not become infected, we must examine the probability that none of the edges of a degree $k$ node transmit their infection. The failure to pass on infection occurs either because the neighbour was itself uninfected with probability $u_\bot$, or that it was infected, but didn't transmit in this instance with probability $(1-u_\bot)(1-T_1)$. Since there are $k$ such edges, and they are all independent of one another, the probability that a node fails to become infected at all is simply the sum of these two probabilities raised to the power $k$
\begin{equation}
    [1-T+u_\bot T_1]^k\nonumber
\end{equation}
The degree is then averaged over the degree distribution of the network to obtain the total probability that the average node remains uninfected as 
\begin{equation}
    \sum_{k=0}^\infty p(k) [1-T_1+u_\bot T_1]^k\nonumber
\end{equation}
At the binary state equilibrium point of bond percolation, nodes are either infected or uninfected. Familiar generating function theory utilises this mutually exclusive relationship to calculate the outbreak fraction of strain 1 as 1 minus the fraction of uninfected nodes. 

\begin{figure*}[ht!]
\begin{center}
\includegraphics[width=0.98\textwidth]{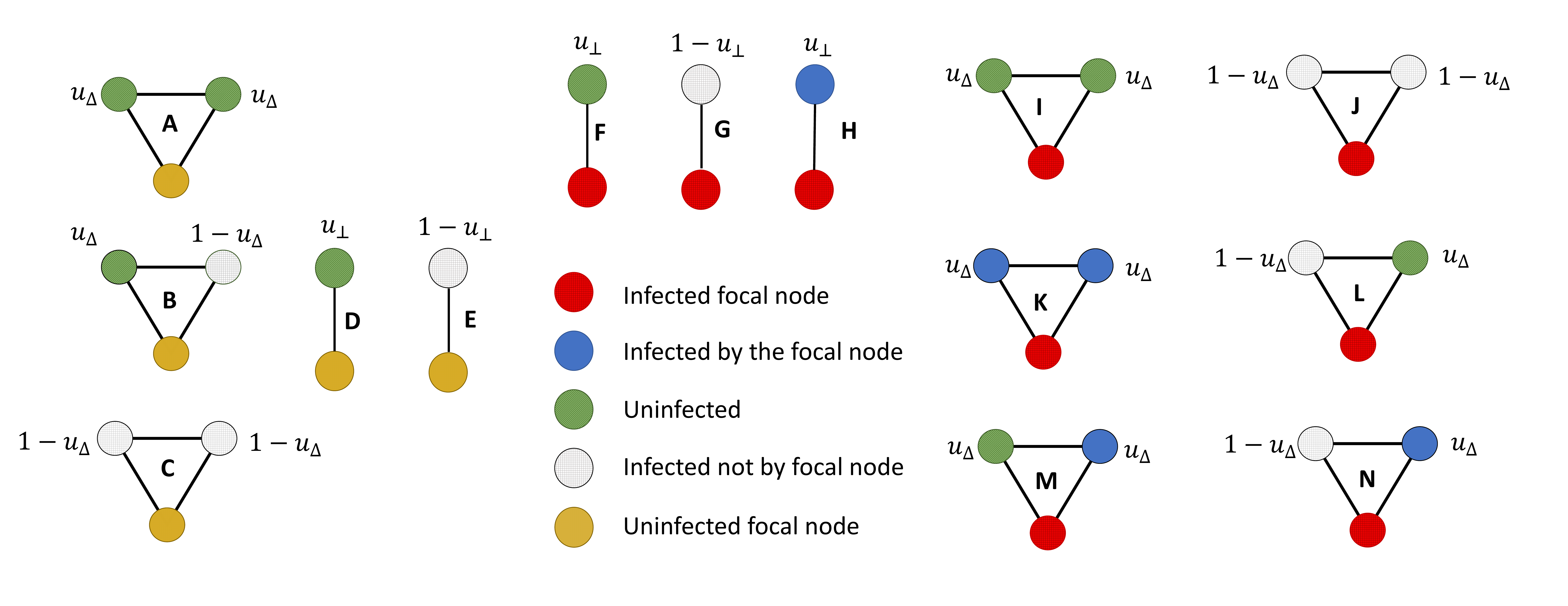}
\caption[trianglesandtrees]{ 
The 14 motifs that surround both focal nodes in the percolation model. In each case the lowest node in the motif is the focal node; of which there are two considerations, uninfected (yellow) and infected (red). There are three states a neighbouring node could be in: uninfected (green), infected externally (grey) and infected directly (blue), although the latter only occurs when the focal node is infected. Motifs [A-C] are the triangle motifs surrounding an uninfected focal node; D and E are the two types of tree-like edges. Motifs [F-H] are the tree-like edges that could surround an infected focal node and finally, motifs [I-N] are the triangles that an infected focal node can belong to. Among these motifs, there are 18 unique node-sites in total. Symmetric triangles (about a vertical axis bisecting the focal node) only contribute one site-type, whilst mixed triangles contribute two site-types and tree-like edges contribute one site-type each. The numbering convention for mixed triangles always proceed from left to right; for instance, in the mixed state triangle B the uninfected neighbour is B1 whilst the infected neighbour is B2. Also, we do not have to include the mirror image of mixed triangles, since, they occur with equal probability. 
} \label{fig:trianglesandtrees}
\end{center}
\end{figure*}

This picture was expanded upon by Newman \textit{et al} \cite{newman_2005, 10.1371/journal.pone.0071321} to describe the equilibrated network from the perspective of both an uninfected node and an infected node \textit{given} they belong to the RG and the GCC, respectively. It is within this picture that we can create a partial immunity model by an extension and adaptation of Newman's work. In a partial immunity model, we cannot simply select a single node-type to describe the final state of the network, and use the mutually exclusive property as before. This is because, subsequent strains can spread on both neighbour types, and so neglecting one of the descriptions leads to an under counting of the full spectrum of transmission routes to the average node in the network. Further, each distinct transmission route from neighbours that have different infection histories to one another occurs with a different probability. Thus, we must describe the equilibrium of strain 1 by considering the local structure of both possible node-states as
\begin{equation}
    1 = p_{\text{uninfected}} + p_{\text{infected}}\label{eq:state_normal}
\end{equation}
In the following two sections, we  describe the local environment of a node in the RG and a node in the GCC following the first percolation process. During the discussion, we refer to the graph motifs in Figure \ref{fig:trianglesandtrees} where each possible neighbour state following percolation of a clustered network is displayed. 

Across both focal nodes, we observe 14 different motifs that could surround a pair of nodes chosen at random. Among these, there are 18 different neighbour states with unique infection histories comprised from three basis states of neighbour node: uninfected (green), infected externally (grey) and infected directly (blue). For example, consider the infected focal node (red) in Figure \ref{fig:trianglesandtrees}. There are 9 different motifs [F-N] that could potentially surround the focal node. Counting each tree-like neighbour and each node within a triangle (excluding the focal node itself) that is not related by symmetry to its neighbour, there are 12 different neighbouring sites; each site is occupied by one of three infection states: uninfected (green), externally infected (grey) and directly infected (blue).

\begin{widetext}
\subsection{Uninfected node description}
The local environment of a node in the RG created by the first strain is considered here. This accounts for all motifs that have the yellow focal node in Figure \ref{fig:trianglesandtrees}. This result was first derived in \cite{PhysRevE.103.062308} and so we present and discuss the result rather than re-derive it. The generating function for the probability of choosing an uninfected focal node (yellow) from the network, $p_\text{uninfected}=F_0(\vec x)$, is 
\begin{align}
F_0(\vec x) =& G_0(u_\bot x_1+(1-u_\bot)(1-T_1)x_2, (u_\Delta x_3)^2+ ((1-u_\Delta)(1-T_1)x_4)^2 + 2u_\Delta(1-u_\Delta)(1-T_1)(1-T_1^2)x_5)\label{eq:F0}
\end{align}
where $u_\Delta$ is the probability that a neighbour node in a triangle is uninfected.
The vector $\vec x = \{x_1,\dots,x_5\}$ has 5 dimensions, one for each of the 5 motifs that could surround the uninfected focal node. We define two additional generating functions $F_{1,\bot}(\vec x)$ and $F_{1,\Delta}(\vec x)$ by replacing $G_0(\vec x)$ in Eq (\ref{eq:F0}) with $G_{1,\bot}(\vec x)$ and $G_{1,\Delta}(\vec x)$, respectively. It is well-known that Eq (\ref{eq:F0}) can be used to generate the size of the GCC of strain 1 according to the following self-consistent set of equations
\begin{align}
    u_\bot =& F_{1,\bot}(\vec 1)\\
    u_\Delta =& F_{1,\Delta}(\vec 1)
\end{align}
followed by $S_1=1-F_{0}(\vec 1)$.

\subsection{Infected node description}

The local environment of a node in the GCC created by the first strain is presented here. This result was first derived in \cite{2020arXiv201209457M} and so for brevity we do not repeat that work. The generating function, $p_\text{infected}=H_0(\vec y)$, for picking an (externally) infected focal node (red) from the network that is part of the GCC is given by
\begin{align}
    H_0(\vec y)=\ &
    G_0(u_\bot(1-T_1)y_1+(1-u_\bot)y_2 + u_\bot T_1y_3,(u_\Delta(1-T_1))^2y_4+(1-u_\Delta)^2y_5+(u_\Delta T_1)^2y_6 \nonumber\\
    &+ 2u_\Delta(1-T_1)(1-u_\Delta)y_7+2u_\Delta(1-T_1)u_\Delta T_1y_8 + 2(1-u_\Delta)u_\Delta T_1y_9)\nonumber\\
    &  - G_0(u_\bot(1-T_1)y_1+(1-u_\bot)(1-T_1)y_2+ u_\bot T_1y_3,(u_\Delta(1-T_1))^2y_4+((1-u_\Delta)(1-T_1))^2y_5+(u_\Delta T_1)^2y_6 \nonumber\\
    & +2u_\Delta(1-T_1)^2(1-u_\Delta)(1-T_1^2)y_7+2u_\Delta(1-T_1)u_\Delta T_1y_8 + 2(1-u_\Delta)u_\Delta T_1(1-T_1)(1-T_1^2)y_9)\label{eq:H0}
\end{align}
We will also define $H_{1,\bot}(\vec y)$ and $H_{1,\Delta}(\vec y)$ by replacing $G_0(\vec y)$ by $G_{1,\bot}(\vec y)$ and $G_{1,\Delta}(\vec y)$, respectively. Additionally, we generate a description of the directly infected neighbour state (blue) as 
\begin{align}
    J_{1,\tau}(\vec y) =\ & G_{1,\tau}(u_\bot(1-T_1)y_1+(1-u_\bot)(1-T_1)y_2+ u_\bot T_1y_3,(u_\Delta(1-T_1))^2y_4+((1-u_\Delta)(1-T_1))^2y_5+(u_\Delta T_1)^2y_6\nonumber\\
    &+2u_\Delta(1-T_1)^2(1-u_\Delta)(1-T_1^2)y_7+2u_\Delta(1-T_1)u_\Delta T_1y_8 + 2(1-u_\Delta)u_\Delta T_1(1-T_1)(1-T_1^2)y_9)
\end{align}
\end{widetext}

The size of the GCC of strain 1 can be found by solving
\begin{subequations}
\begin{align}
    u_\bot =& J_{1,\bot} (\vec 1)\\
    u_\Delta = & J_{1,\Delta}(\vec 1)\label{eq:u1vales}
\end{align}
\end{subequations}
and then $S_1=H_{0}(\vec 1)$. In relation to the uninfected node description we have that $F_{1,\tau}(\vec 1) = J_{1,\tau}(\vec 1)$ and that $H_0(\vec 1) = 1-F_0(\vec 1)$. Thus, the full description of the binary state equilibrium following bond percolation is given by the relation  
\begin{equation}
    1 = F_0(\vec 1) + G_0(\vec 1)
\end{equation}
This expression constitutes a novel way, to our knowledge, of using the generating function formulation and it is this key equation that allows us to create the partial immunity model. 

\section{Strain 2}

We have seen above how the GCC of the first strain can be obtained from either description of members of the percolation equilibrium: an uninfected node in the RG or an infected node in the GCC. Both methods utilise the state normalisation in Eq (\ref{eq:state_normal}) and the mutually exclusive property of the binary state. 

To calculate the outbreak size of strain 2, we proceed as follows. For each of the 18 possible neighbouring node states, we must introduce a probability that infection with strain 2 does not occur through this channel by some means. Therefore, we introduce 18 distinct probabilities that a neighbour of a given state fails to infect a given focal node with strain 2. Although arbitrary, we choose different symbols for these probabilities depending on whether the neighbour state surrounds an uninfected node or an infected node. We will see in a moment that subsets of the 18 sites are generated by the same expressions, and as such, the dimensionality of the model can be significantly reduced. However, we proceed in full for the moment.  

There are 6 unique states that surround an uninfected focal node and thus, we define a set of 6 probabilities, $\{w\}$, that each hold the value of not becoming infected by strain 2 from one of these states. Specifically, there are 4 triangle neighbours and 2 tree-like neighbours so 
\begin{equation}
    \{w\} = \{w_\Delta^A, w_\Delta^{B1},w_\Delta^{B2},w_\Delta^C, w_\bot^D,w_\bot^E\}
\end{equation}
Similarly, there are 12 states surrounding the node in the GCC and so we introduce a set, $\{v\}$, that holds the values of the probabilities of not becoming infected with strain 2 from these states. Specifically, there are three states reached by tree-like edges and 9 states within the triangle motifs. Hence, 
\begin{align}
    \{v\} =&\  \{ v_\bot^F,v_\bot^G,v_\bot^H,v_\Delta^I,v_\Delta^J,v_\Delta^K,v_\Delta^{L1},v_\Delta^{L2},v_\Delta^{M1},\nonumber\\
    &\ v_\Delta^{M2},v_{\Delta}^{N1},v_\Delta^{N2}\}
\end{align}

We next need to write self-consistent expressions for each of the values in $\{w\}$ and $\{v\}$. Before we do this, we define two functions that express the probability of transmission failing through a tree-like edge, $g(v,T) = v +(1-v)(1-T)$, and a triangle motif
\begin{align}
    h(v_\mu,v_\nu,T_\mu,T_\nu) =\ & g(v_\mu,T_\mu)g(v_\nu,T_\nu)\nonumber\\
    &- v_\mu(1-v_\nu)(1-T_\nu)T_\nu T_\mu\nonumber\\
    & - v_\nu(1-v_\mu)(1-T_\mu)T_\mu T_\nu
\end{align}
with the convention that $h(v_\mu,v_\mu,T_\mu,T_\mu) = h(v_\mu,T_\mu)$. We will insert these functions into the $\vec x $ and $\vec y$ vectors in the arguments of the generating functions; each insertion describing the probability that strain 2 isn't contracted from a particular motif. The probability of not getting infected by strain 2 from the uninfected neighbour at the end of a tree-like edge is 
\begin{subequations}
\begin{align}
w_\bot^D = F_{1,\bot}/u_\bot
\end{align}
The probability of not contracting strain 2 from the infected neighbour at the end of a tree-like edge is 
\begin{align}
w_\bot^E =&\ H_{1,\bot}/(1-u_\bot)
\end{align}

We now turn to the triangle probabilities $\{w_\Delta^A, w_\Delta^{B1},w_\Delta^{B2},w_\Delta^C\}$. The probability that the uninfected focal node doesn't get strain 2 from the symmetric susceptible site is 
\begin{align}
w_\Delta^A = F_{1,\Delta}/u_{\Delta}
\end{align}
The probability that the symmetric infected site doesn't transmit to the uninfected focal node is 
\begin{align}
w_\Delta^C =&\ H_{1,\Delta} /(1-u_\Delta)
\end{align}
The mixed triangle follows. For the uninfected focal node, we have the probability of not becoming infected with strain 2 from an uninfected neighbour as 
\begin{align}
w_\Delta^{B1} = F_{1,\Delta} /u_{\Delta}
\end{align}
Whilst for the infected site we have 
\begin{align}
w_\Delta^{B2} =&\ H_{1,\Delta} /(1-u_\Delta)
\end{align}
We now have all of the probabilities that we require to describe the local environment of the uninfected node. We now turn to the description of the infected node in the GCC of strain 1. The three tree-like sites, $\{ v_\bot^F,v_\bot^G,v_\bot^H\}$, are generated as follows: the uninfected neighbour 
\begin{align}
    v_\bot^F =&\ F_{1,\bot}/u_\bot
\end{align}
the externally infected neighbour
\begin{align}
    v_\bot^G =&\ H_{1,\bot}/(1-u_\bot)
\end{align}
and the directly infected neighbour
\begin{align}
v_\bot^{H} =&\  J_{1,\bot}/u_\bot
\end{align}
We now require the 9 triangle values $\{v_\Delta^I,v_\Delta^J,v_\Delta^K,v_\Delta^{L1},v_\Delta^{L2},v_\Delta^{M1},v_\Delta^{M2},v_{\Delta}^{N1},v_\Delta^{N2}\}$. The probability that an uninfected neighbour fails to transmit strain 2 through a symmetric uninfected triangle I is
\begin{align}
v_\Delta^I = F_{1,\Delta}/u_\Delta
\end{align}
The probability that the infected focal node in triangles J and K does not contract strain 2 is 
\begin{align}
    v_\Delta^J =&\ H_{1,\Delta}/(1-u_\Delta)
\end{align}
and
\begin{align}
    v_\Delta^K =&\ J_{1,\Delta}/u_\Delta
\end{align}
The mixed triangle L is given by 
\begin{align}
    v_\Delta^{L_1} =& H_{1,\Delta}/(1-u_\Delta)
\end{align}
and 
\begin{align}
v_\Delta^{L2} = F_{1,\Delta} /u_\Delta
\end{align}
Triangle M follows as 
\begin{align}
v_\Delta^{M1} = F_{1,\Delta} /u_\Delta
\end{align}
and
\begin{align}
    v_\Delta^{M_2} =&\ J_{1,\Delta}/u_\Delta
\end{align}
Finally, triangle N is given by 
\begin{align}
    v_\Delta^{N_1} =& H_{1,\Delta}/(1-u_\Delta)
\end{align}
and
\begin{align}
    v_\Delta^{N_2} =&\ J_{1,\Delta}/u_\Delta
\end{align}
\end{subequations}
At this point, we have not yet written the arguments of each generating function, $\vec x$ and $\vec y$. It happens, that there are several equivalent expressions among the variables, allowing us to reduce the dimension of the problem considerably. Specifically, we notice the following redundancies among the relations:
$
    v_\Delta^{M1} = v_\Delta^{L2} = v_\Delta^I= w_\Delta^{B1}=w_\Delta^A
$, 
$
    w_\Delta^{B2} = w_\Delta^C = v_\Delta^{L1} = v_\Delta^{N1} = v_\Delta^{J}
$, $
w_\bot^E=v_\bot^G
$, 
$
    w_\bot^D = v_\bot^F
$, and 
$
    v_\Delta^K=v_\Delta^{N2}=v_\Delta^{M2}
$.
This over prescription affords a reduction in the number of system variables to only 6 independent variables, one for each of the possible neighbour nodes: uninfected, externally infected and directly infected for tree-like edges and triangle motifs, respectively. Therefore, if we write the argument of each generating function $F_{1,\bot},H_{1,\bot},J_{1,\bot},F_{1,\Delta},H_{1,\Delta}$ and $J_{1,\Delta}$ once, it is known for all occurrences of that function in the model. Further, we observe that the only difference between  $F_{1,\tau}, H_{1,\tau}$ and $J_{1,\tau}$ for $\tau=\bot,\Delta$ is the underlying $G_{1,\tau}$ function, not the argument. In other words, the arguments of $F_{1,\bot}$ and $F_{1,\Delta}$, for instance, are equivalent; we do not distinguish based on their topology. A final simplification can be achieved by noting that the arguments of $J_{1,\tau}$ and $H_{1,\tau}$ are also equivalent. Therefore, there are only two arguments to write: one for $F_{1,\tau}$ and another for $H_{1,\tau}$. These are given by $\vec x=\vec\zeta$ and $\vec y = \vec\xi$ where
\begin{widetext}
\begin{align}
\vec \zeta = \{g(w_\bot^D,T_2),g(w_\bot^E,T_2'),h(w_\Delta^A,T_2),h(w_\Delta^C,T_2'),h(w_\Delta^{B1},w_\Delta^{B2},T_2,T_2')\}
\end{align}
and
\begin{align}
\vec\xi =\ &\{g(v_\bot^F,T_2),g(v_\bot^G,T_2'),g(v_\bot^H,T_2'),h(v_\Delta^I,T_2),h(v_\Delta^J,T_2'),h(v_\Delta^K,T_2'),h(v_\Delta^{L1},v_\Delta^{L_2},T_2',T_2),\nonumber\\
&\times h(v_\Delta^{M1},v_\Delta^{M2},T_2,T_2'),h(v_\Delta^{N_1},v_\Delta^{N_2},T_2',T_2')\}
\end{align}
which constitute vectors of probabilities that each neighbour site fails to transmit infection to the focal node (or connect it to the GCC). With this in place, we now have an expression for all of the required probabilities $\{w\}$ and $\{v\}$. The size of the second outbreak over the network is then found by solving
\begin{align}
    S_2 =& [ F_{0}(\vec 1)+ H_{0}(\vec 1)] -[ F_{0}(\vec \zeta)+ H_{0}(\vec \xi)]\label{eq:S2}
\end{align}
where $[ F_{0}(\vec 1)+ H_{0}(\vec 1)]= 1$. Qualitatively, this expression is 1 minus the probability that a node obtains strain 2 from either uninfected or infected neighbours. 
\begin{figure*}[ht!]
\begin{center}
\includegraphics[width=1\textwidth]{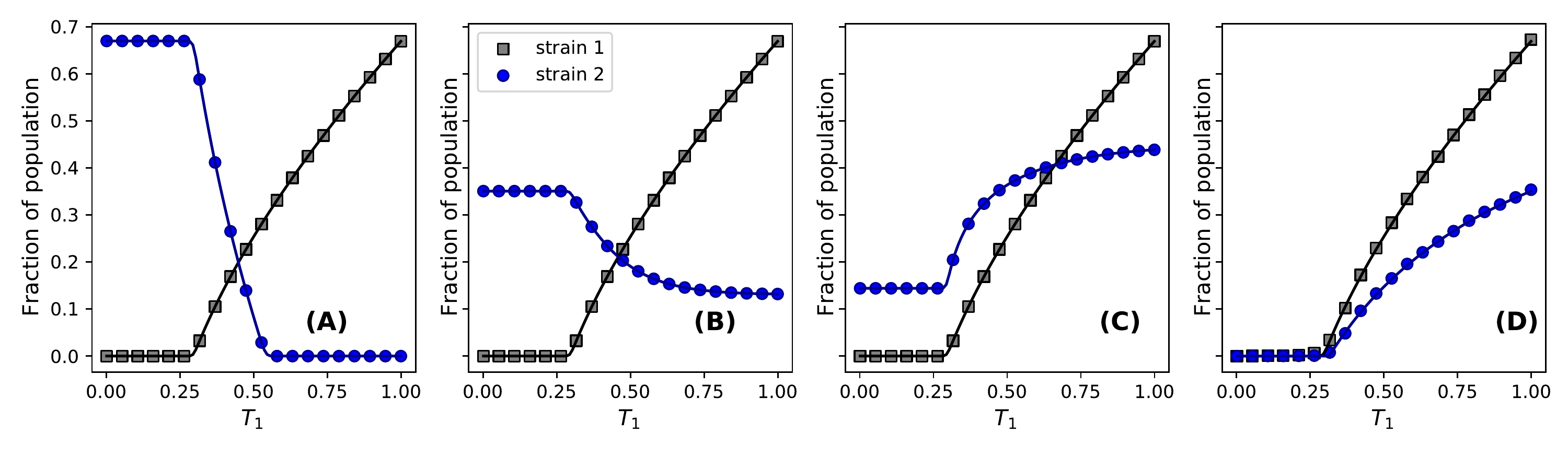}
\caption[maintrees]{ 
The outbreak fractions for several $(T_2,T_2')$ combinations of the model all in the absence of clustering. From left to right: (A) complete cross-immunity $(T_2,T_2')=(1,0)$ , (B) partial cross-immunity $(T_2,T_2')=(0.6,0.39)$ , (C) partial coinfection $(T_2,T_2')=(0.4,0.7)$ and (D) perfect coinfection $(T_2,T_2')=(0,0.6)$ . Markers are the average of 50 repeats of bond percolation over CCM networks of size $N=65000$, $\alpha=2.0$ and $\theta=0$; square markers are strain 1 whilst circles are strain 2. Solid lines are the theoretical results of Eq (\ref{eq:S2}) for strain 2. 
} \label{fig:maintrees}
\end{center}
\end{figure*}

\end{widetext}

\section{Numerical Results}

The results of the model under 4 different strain interactions for tree-like networks in the absence of clustering are shown in Figure \ref{fig:maintrees} as $T_1$ is varied. Across the simulations, the networks are built according to the clustered contact model, (CCM), defined in \cite{2020arXiv201209457M}, which is an example of a configuration model degree distribution with clustering. Power law contact distributions are typical of those found in real-world social networks \cite{newman_20055}. The underlying degree distribution is given by a power law model with exponential degree cut-off (PLC) defined as
\begin{equation}
    p^{\text{PLC}}(k) = \frac{k^{-\alpha}e^{-k/\kappa}}{\text{Li}_\alpha(e^{-1/\kappa})}
\end{equation}
where $\kappa$ is the degree cut-off, $\alpha\in [2,3]$ is a power law exponent and $\text{Li}_n(z)$ is the $n$th polylogarithm of $z$ \cite{PhysRevE.66.016128}. Each $k$ is then decomposed into tree-degrees, $s$ and triangle degrees, $t$, according to
\begin{equation}
    p^{\text{CCM}}(k) = p^{\text{PLC}}(k) \sum_{t=0}^{\lfloor k/2\rfloor}\binom{\lfloor k/2\rfloor}{t}\theta^{t}(1-\theta)^{\lfloor k/2\rfloor-t}\label{eq:CHCN}
\end{equation}
where $\lfloor \cdot \rfloor$ is the floor function and $\theta\in [0,1]$ is the probability of a pair of edges belonging to a triangle. 

We simulate bond percolation for both
strains numerically using Monte Carlo simulations. Following strain 1, infected nodes are labelled and subsequent infection with strain 2 occurs with probability $T_2$ for nodes in the RG or $T_2'$ for GCC nodes. 

In Figure \ref{fig:maintrees}a we have $T_2=1$ and $T_2'=0$, a perfect cross-immunity model \cite{newman_2005} in which infection with strain 1 prevents infection with strain 2. In Figure \ref{fig:maintrees}b we relax this hard limit, with $T_2=0.6$ and $T_2'=0.39$, to obtain a partially cross-immune interaction whereby the transmission of strain 2 is reduced for strain 1 infected nodes. For $T_1<T_{1,c}$ we observe the steady-state of strain 2 without competition from strain 1. At $T_1=T_{1,c}$ a GCC in strain 1 emerges and the number of cases of strain 2 drops, but does not vanish; in the limit $T_1=1$ strain 2 reaches its lowest incidence rate as competition is maximised. In Figure \ref{fig:maintrees}c we observe a partial coinfection model, with $T_2=0.4$ and $T_2'=0.7$. In this case, strain 2 is facilitated by the presence of strain 1 in the network; the symbiotic interaction leading to an increase in the incidence of strain 2 infected nodes. Figure \ref{fig:maintrees}d shows the hard limit of a perfect coinfection model \cite{10.1371/journal.pone.0071321} with $T_2=0$ and $T_2'=0.6$, strain 2 cannot survive without a GCC of strain 1 present in the network. 
\begin{figure}[ht!]
\begin{center}
\includegraphics[width=0.38\textwidth]{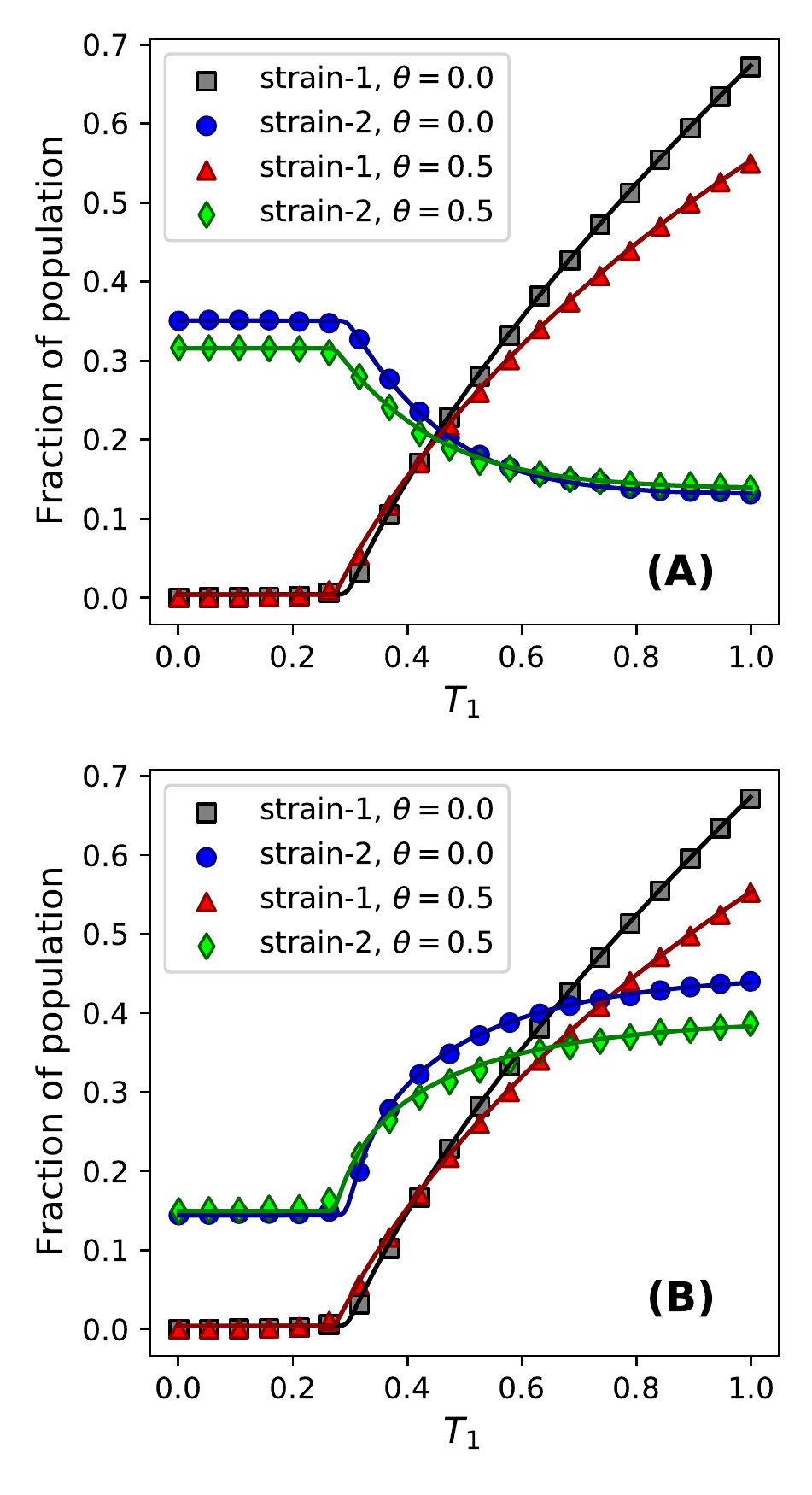}
\caption[maintriangle]{ 
The outbreak fractions for both strains for clustered (Eq \ref{eq:CHCN}) and unclustered networks as $T_1$ is varied under two disease couplings: (A) partial cross-immunity $(T_2,T_2')=(0.6,0.4)$ and (B) partial coinfection $(T_2,T_2')=(0.4,0.7)$ . Simulations are the average of 50 repeats of bond percolation on networks with $N=35000$ and $\theta=0.0$, $\alpha=2.0$ and $0.5$ for the unclustered and clustered networks, respectively. Solid lines are the theoretical results of Eqs \ref{eq:H0} and \ref{eq:S2}. In general, clustering reduces the extent of plural infections in the network; however, degree assortativity within the contact topology causes a reversal of this at high (low) values of $T_1$ in A (B). 
} \label{fig:maintriangles}
\end{center}
\end{figure}

With an understanding of the model without clustering, we now examine the case where $\theta\neq 0$ for both partial interaction models with $\kappa=20$, $\alpha=2$ and $\theta = 0.5$, see Figure \ref{fig:maintriangles}. The epidemic threshold of strain 1 is reduced with clustering, so too is the overall outbreak size of strain 1 at large $T_1$, compared to unclustered networks. The incidence of strain 2 exhibits dual behaviour over the range of $T_1$. For the partial immunity scenario (Figure \ref{fig:maintriangles}a), with $T_2=0.6$ and $T_2'=0.4$, clustering reduces the incidence of strain 2 at low $T_1$; however, it increases it as $T_1\rightarrow 1$. Conversely, for partial coinfection (Figure \ref{fig:maintriangles}b), with $T_2=0.4$ and $T_2'=0.7$, having lowered the epidemic threshold of strain 1, clustering causes an increase in the incidence of strain 2 at lower $T_1$ values compared to the unclustered analogue.

\begin{figure}[ht!]
\begin{center}
\includegraphics[width=0.38\textwidth]{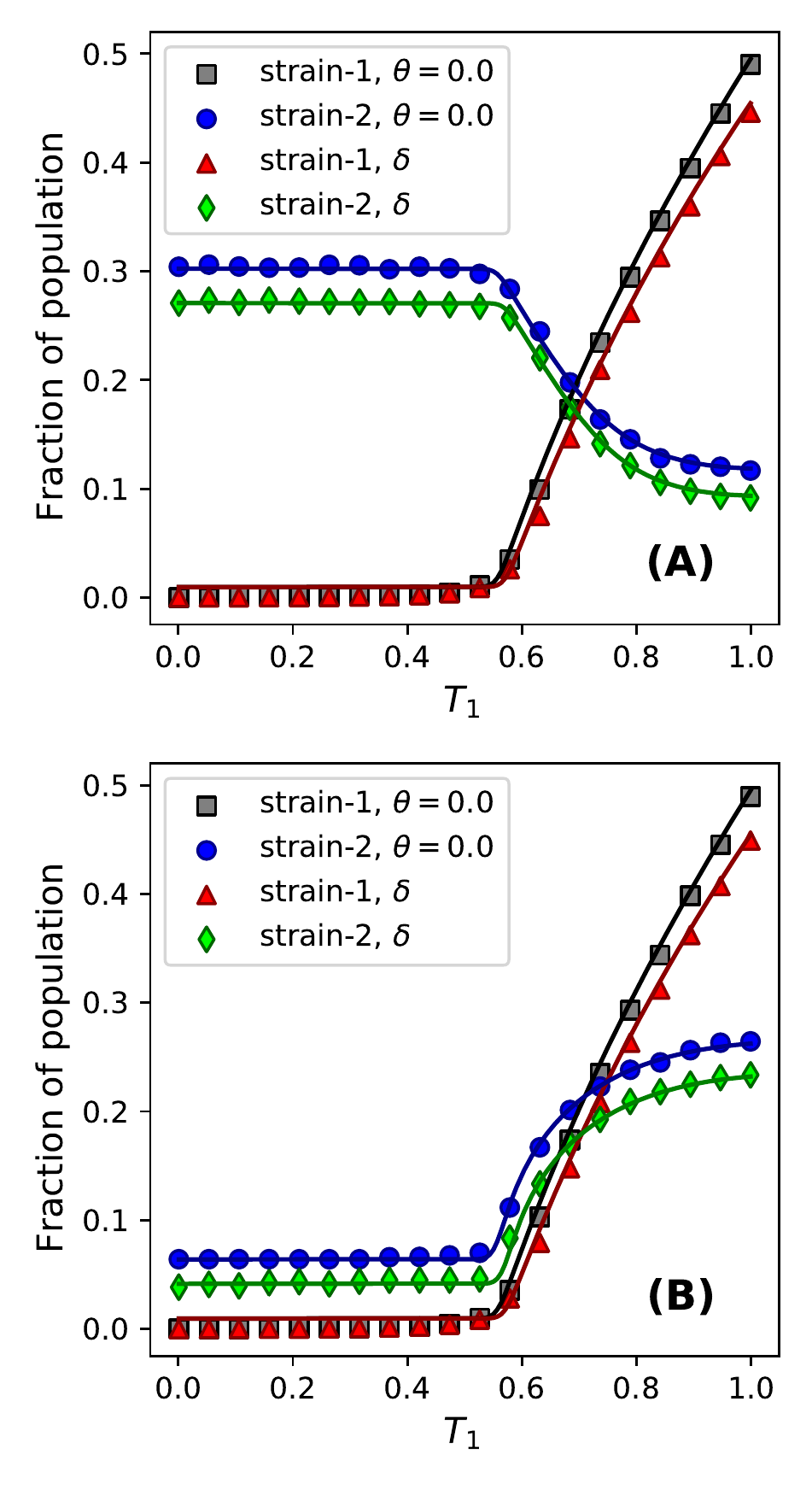}
\caption[deltatriangle]{ 
The outbreak fractions of the degree-$\delta$ model with clustering constrained to low degree assortativity for (A) partial immunity $(T_2,T_2')=(0.8,0.65)$  and (B) partial coinfection $(T_2,T_2')=(0.6,0.75)$ disease interactions. Simulations are the average of 50 repeats of bond percolation on networks with $N=35000$, $\alpha=2.0$; solid lines are the theoretical results of Eqs \ref{eq:H0} and \ref{eq:S2}. Clustering reduces the fraction of the network that becomes infected by both strains for all values of $T_1$.
} \label{fig:delta}
\end{center}
\end{figure}

In Figure \ref{fig:delta} we perform a second experiment using the degree-$\delta$ model \cite{miller_2009,PhysRevE.81.066114,2020arXiv201209457M}. We define a distribution in which the degree of nodes involved in triangles is fixed to $k=3$ and thus $(s,t)=(1,1)$, whilst all other degrees are given by Eq (\ref{eq:CHCN}) for $(s,t)=(k,0)$.
With the degree-correlations among triangles fixed, the epidemic threshold of the first strain increases with clustering. The partial cross-immune coupling (Figure \ref{fig:delta}a), with $T_2=0.8$ and $T_2'=0.65$, no longer exhibits a cross-over in expected size of strain 1 and strain 2; clustering reduces the incidence of strain 2 for all values of $T_1$. Similarly, the partial coinfection model (Figure \ref{fig:delta}b), with $T_2=0.6$ and $T_2'=0.75$, exhibits a reduced incidence of strain 2 compared to the unclustered analogue. As $T_1\rightarrow 1$, however, the coinfection is reduced in the clustered graph compared to the unclustered.

\section{conclusion}

In this paper we have presented a general model of two pathogens spreading sequentially over a clustered network. The two strains can interact in a very broad manner ranging from perfect coinfection, facilitating the spreading of the subsequent strain, to perfect antagonism, competing for hosts, and all combinations in between. Our model uses generating functions to provide a theoretical understanding to Monte Carlo simulations of bond percolation, showing excellent agreement. Our model, which considers the local description of both uninfected and infected nodes simultaneously, introduces a novel way to use generating functions to describe bond percolation. The paradigm of describing the entire absorbing state of the percolation equilibrium rather than the typical approach of only the uninfected nodes will certainly prove illuminating to many areas of epidemic spreading and network dynamics.  

Our simulations on clustered networks examined how the dynamics of strain 2, under a partial disease interaction, were influenced by contact topology. We found that clustering generally reduces the prevalence of strain 2 in the network. The magnitude of this effect being dependent on the details of the clustering and the value of the transmissibility of strain 1. However, networks with finite degree correlations present (due to contact clustering) were found to have greater incidences of coinfection.

It is clear that this model can be reduced to purely tree-like networks by the removal of references to triangle motifs in the generating functions. The work herein can be generalised in a number of ways: firstly, we have formulated a clear recipe to follow for the inclusion of different types of clustering, such as higher-order cliques or cycles \cite{2020arXiv200606744M} or indeed custom motifs such as \cite{mann2020random}. In many cases, these subgraphs may represent social networks more accurately than the tree-triangle model we have considered here, and thus, this extension is important for the rationalisation of disease spreading among human populations.

As presented, the model considers sequential strains that are temporally separated. Often, diseases spread simultaneously among a population. It is well-known from analytical methods based on differential equations \cite{PhysRevX.4.041005,10.1371/journal.pcbi.1002042} and simulations of SIR through, for instance, Gillespie simulation, that the initial conditions of each stain/disease greatly influence the dominant pathogen for two otherwise symmetric diseases \cite{PhysRevE.84.036106}. Bond percolation and the method of generating functions inherently cannot capture the stochastic effects of concurrent spreading due to its equilibrium-based nature. However, previous work has been conducted \cite{PhysRevE.84.036106,PhysRevE.81.036118} to model a mutual pathogen interaction using generating functions. Extending our model to the study of concurrent strains would be a significant step forward for understanding disease spreading. 

The model could be applied to multilayer clustered networks with some adaptation to increase the influence of topology on the spreading and enrich the dynamics further. Or perhaps the effects of drugs or vaccines that target particular topologies or node sites could also be investigated.

However, perhaps the most significant future generalisation of this model is the extension to additional strains, in other words, subsequent percolations. As we do this, the immunological landscape becomes increasingly rich and the node's infection histories more diverse. We believe that the best approach would be to develop an automation, perhaps through recursion, that can write the required expressions, in a similar vein to \cite{mann2020random}. This would allow the exact study of an $N$-strain seasonal influenza model and could also be used to discover once-a-century events (such as the 1918-1919 Spanish influenza or the COVID-19 pandemic), along with generational evolutionary pressures and genetic drift. We also believe that the parameters controlling disease interactions could be matched to real-world data to understand the relationship, symbiotic or antagonistic, between multiple diseases as they evolve and spread.

\subsection*{References}

\bibliography{bib}

\end{document}